\documentclass[twocolumn,preprintnumbers,amsmath,amssymb]{revtex4}
\input{epsf}

\newcommand{\be}{\begin{equation}}
\newcommand{\ee}{\end{equation}}
\newcommand{\bea}{\begin{eqnarray}}
\newcommand{\eea}{\end{eqnarray}}
\newcommand{\bef}{\begin{figure}}
\newcommand{\eef}{\end{figure}}

\renewcommand{\phi}{\varphi}
\renewcommand{\rho}{\varrho}

\newcommand{\simge}{\gtrsim}
\newcommand{\simle}{\lesssim}

\newcommand{\eV}{\mbox{eV}}

\newcommand{\MeV}{\mbox{MeV}}
\newcommand{\GeV}{\mbox{GeV}}

\newcommand{\pc}{\mbox{pc}}
\newcommand{\kpc}{\mbox{kpc}}
\newcommand{\Mpc}{\mbox{Mpc}}
\newcommand{\G}{\mbox{G}}

\newcommand{\Rey}{{R_e}}

\newcommand{\vA}{{\bf A}}
\newcommand{\vB}{{\bf B}}

\newcommand{\He}{{\cal H}}

\def\eps@scaling{0.96}

\def\showone#1{
  \centering
  \leavevmode
  \epsfxsize=\eps@scaling\linewidth
  \epsfbox{#1.eps}
}

\def\epstwo@scaling{0.48}

\def\showtwo#1#2{
  \centering
  \leavevmode
  \epsfxsize=\epstwo@scaling\linewidth
  \epsfbox{#1.eps} \hfil
  \epsfxsize=\epstwo@scaling\linewidth
  \epsfbox{#2.eps}
}

\begin{document}

\title{Are Cluster Magnetic Fields Primordial ?}
\author{Robi Banerjee$^1$ and Karsten Jedamzik$^{2}$}
\affiliation{$^1$Department of Physics and Astronomy, McMaster
  University, Hamilton, ON L8S 4M1, Canada
} 
\affiliation{$^2$Laboratoire de Physique Math\'emathique et Th\'eorique, C.N.R.S.,
Universit\'e de Montpellier II, 34095 Montpellier Cedex 5, France}

\begin{abstract}
We present results of a detailed and fully non-linear numerical and 
analytical investigation of magnetic field evolution from the very 
earliest cosmic epochs to the present. 
We find that, under reasonable assumptions concerning the efficiency
of a putative magnetogenesis era during cosmic phase transitions,
surprisingly strong magnetic fields $10^{-13} - 10^{-11}\,\G$, on
comparatively small scales $100\,\,\pc$ -- $10\,\,\kpc$ may survive 
to the present.
Building on prior work on the evolution of magnetic fields during the
course of gravitational collapse of a cluster, which indicates
that pre-collapse
fields of $\sim 4\times 10^{-12}\,\G$ extant on small scales may 
suffice to produce
clusters with acceptable Faraday rotation measures, we question the
widely hold view that cluster magnetic fields may not be entirely of
primordial origin. 
\end{abstract}


\maketitle

Magnetic fields exist throughout the observable Universe. They exist
in the interstellar medium, in galaxies, and clusters of
galaxies (for reviews cf.~\cite{rev1}). 
The origin of galactic- and cluster- magnetic fields is still unknown. 
A plausible, though by far not convincingly established possibility is the
generation of magnetic seed fields and their subsequent amplification
via a galactic dynamo mechanism. 
Seed fields may be due to a variety of processes
(and with a variety of strengths), such as the Biermann battery within
intergalactic shocks \cite{kuls97a}, stellar magnetic fields expelled
in planetary nebulae, or during supernovae explosions,
either into the intragalactic, or in the presence of galactic outflows into
the intergalactic medium~\cite{rees87}, as well as due to quasar outflows of
magnetized plasma~\cite{furl01}. Seed fields may also be of
primordial origin with a multitude of proposed scenarios.
These include generation during first-order phase transitions (e.g.
QCD or electroweak), around cosmic defects, or during an inflationary
epoch (with, nevertheless, extremely small amplitudes).  
For a review of proposed scenarios we refer the reader
to~\cite{gras01}.

The philosophy in prior studies of primordial magnetogenesis is often (but not
always) as follows.
After establishing a battery mechanism (e.g. separation of charges and
production of currents)
and a \lq\lq prescription\rq\rq\, or estimate for the final, non-linearly
evolved magnetic field strength 
(e.g. equipartition of magnetic 
energy with turbulent flows), subsequent evolution is
approximated by simply assuming
frozen-in magnetic field lines into the plasma. Though such an
approximation may be appropriate
on the very largest scales, it should be clear, that this may not be
the case on the ``integral'' or coherence scale of the field. Here, coupling
of the magnetic fields to the gas induces non-linear cascades of energy in
Fourier space. The characteristics of initially created magnetic field are thus
vastly modified during cosmic evolution between the epoch of 
magnetogenesis and the present.

The final step in such studies is then often to determine field strengths
on some prescribed large scale (e.g. $10\,\Mpc$) typically falling in the range 
$10^{-30}\,\G\simle B \simle 10^{-20}\,\G$, inferring that this
may act as seed for a sufficiently efficient dynamo to produce  
galactic- and cluster- magnetic fields of order $10^{-6}\,\G$. 
This is observed in negligence of the fact that
much stronger fields on smaller scales, result not only from a variety of
astrophysical seeds, but from these very same primordial scenarios.
Considering the likelihood of a magnetized plasma in the early Universe, 
it seems important to be able to make predictions on such final remnant 
fields surviving a magnetogenesis scenario to the present epoch, 
particularly also since
such fields may fill voids of galaxies and may potentially 
be observable by upcoming technology~\cite{psl}.

The purpose of this letter is twofold. 
We have attempted to develop a coherent picture of gross
features of non-linear, cosmic MHD evolution of primordial fields, 
including all relevant
dissipative processes, such as viscosity due to diffusing- or free-streaming-
neutrinos and photons, as well as ambipolar damping. 
A subset of the
results of our numerical and analytical analysis
is presented here, whereas details are presented elsewhere~\cite{BJ03}. 
Our study allows us, for the first time, to 
make predictions
for magnetic field energy and coherence length at the present epoch,
for broad ranges of initial magnetic 
configurations, parametrized by spectral index, initial helicity, 
initial energy, as well as era of magnetogenesis.
Second, drawing on
a prior numerical MHD study of the gravitational collapse of a 
cluster of galaxies~\cite{DBL99}, we challenge the often-cited
conclusion that cluster 
magnetic fields may not be entirely of primordial origin. Rather, we
stress that it seems not clear at the moment if such fields on comparatively 
small scales may not, after all, produce cluster magnetic fields as 
widely observed.

In passing we note that
there exists a number of 
analytical~\cite{dimo96,jeda98,subr98,son99,field00,vach01,sigl02} , 
and numerical~\cite{brand96b,chris01}, 
studies on the evolution of non-helical and helical primordial
magnetic fields, which, nevertheless, for one
or the other reason either remain inutile in predicting final field 
properties, or do so only for a specific scenario. The inutility of
results is related to facts such as, the adoption of an evolutionary
model not supported by numerical simulations, an inadequate treatment
of viscosity due to photons, or simply, the analysis being linear in
nature or being performed in Minkowski space and not properly transferred
to the expanding Universe.

The generation of primordial magnetic fields in magnetogenesis scenarios
is generally believed to occur during well-defined periods 
(e.g. QCD-transition). Subsequent evolution of these magnetic fields
should therefore be well approximated by a ``free decay'' without any further
input of kinetic or magnetic energy, i.e. as freely decaying MHD. 
The exceedingly large Prandtl numbers in the early 
Universe allow one to neglect dissipative effects due to finite conductivity. 
Nevertheless, as already stressed in Ref.~\cite{jeda98},
dissipative effects arising from the ``imperfectness'' of the fluid due
to neutrino- and photon- diffusion and free-streaming play an important role
in early MHD evolution. One may further show that for field strength
as considered in this letter the assumption of fluid incompressibility
is appropriate.

To verify theoretical expectations we have performed 
numerical simulations of incompressible, freely decaying, ideal, but 
viscous MHD.
These simulations are performed with the help of a modified
version of the code ZEUS-3D~\cite{stone92,clar94} in a
non-expanding (Minkowski) background and on $128^3$ to $512^3$ grids. 
Modifications lie in the inclusion of 
fluid viscosities~\cite{KJAM}. It is
shown elsewhere~\cite{brand96b,BJ03} 
that conformal- or near-conformal- invariance
of the MHD equations allow for the interpretation of results obtained
with Minkowski metric to results applicable for a Friedman-Robertson-Walker
metric. Results of such simulations, in particular the decay of 
magnetic energy $E_{\rm mag}$ with time, for a variety of physical regimes
are shown in Fig.~\ref{fig:E_mag_by_time}.

\bef
\def\eps@scaling{0.83}
\showone{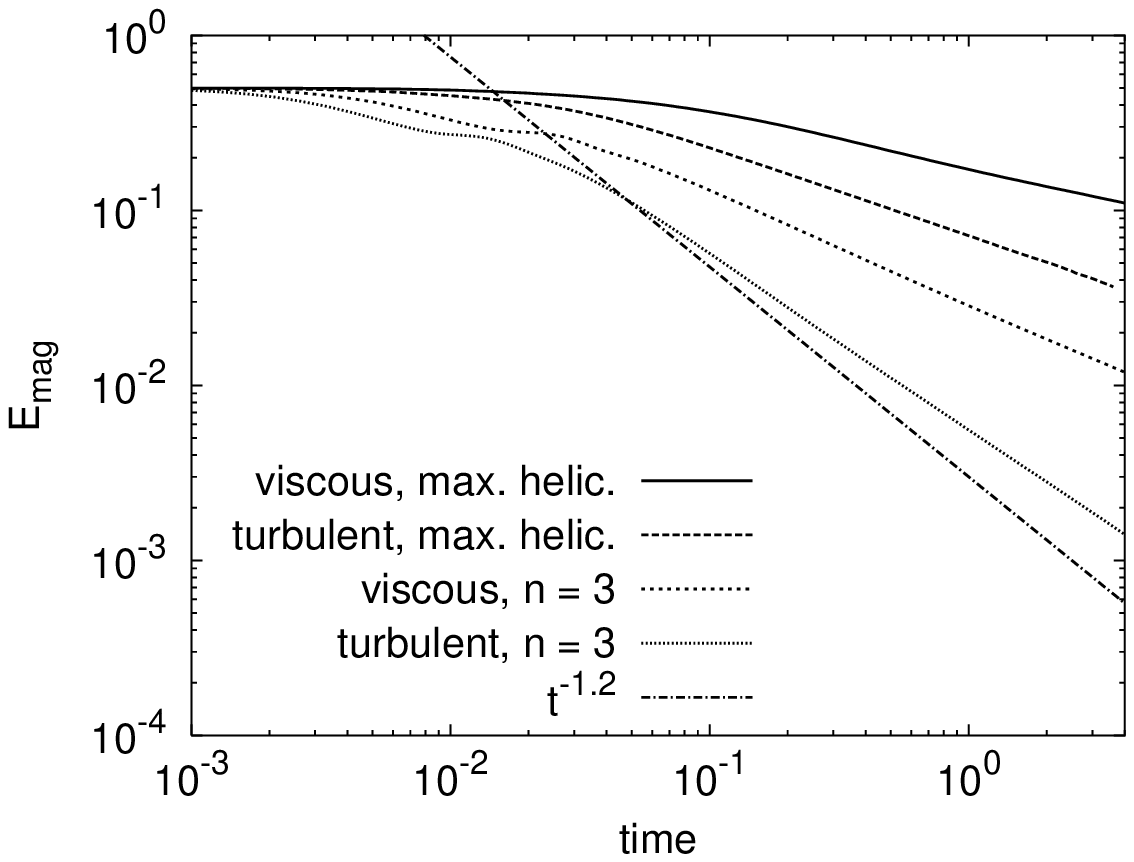}
\caption{Decay of magnetic energy in different damping regimes and for 
different inital conditions as observed in our numerical simulations. The
$t^{-1.2}$ line shows the theoretical damping law in the turbulent
regime for $n=3$, $h_g=0$.} 
\label{fig:E_mag_by_time}
\eef

We have found that results of our simulations may be understood
in a comparatively simple manner. In particular, non-linear
MHD processing of the initial spectrum at epoch with Hubble constant $H(T)$
occurs for all scales $l$, which obey
\begin{equation}
v(l)/l \simge H(T)\,\, ,\label{relax}
\end{equation} 
irrespective of the Reynolds number $\Rey$ of the flow. Here
$v(l)$ may be written as the Alfv\'en velocity
$v_A(l) = B(l)/\sqrt{4\pi(\rho +p)}$ when
turbulence holds, $\Rey\simge 1$, and as $v(l) =
v_A(l)\,\Rey(v=v_A,l)$ for 
viscous MHD ($\Rey\simle 1$). This holds equally during
the photon diffusion ($l_{\gamma}\ll l$) regime 
$\Rey = vl/\eta$ and the photon free-streaming ($l_{\gamma}\gg l$)
regime $\Rey = v/\alpha l$, where $\eta$, $\alpha$, and $l_{\gamma}$ 
are photon shear viscosity, drag coefficient, and mean free path, 
respectively.
Defining $L(T)$, the integral- or coherence- scale, 
as the scale where equality applies in Eq.~(\ref{relax}), one finds  
that $L(T)$ is in fact, the scale containing most of the energy of the flow.
This is due to a non-linear and rapid cascade developing on
all scales $l\simle L$, with energy in fluid eddies transported down to
the dissipation scale $l_{\rm diss}$ and transferred to heat. Since the
resultant small-scale spectrum is red and we assume the initial as yet
unprocessed 
large-scale ($l\simge L$) spectrum to be blue, $L(T)$, as the
smallest unprocessed scale, remains as the
magnetic coherence scale of the field at epoch with temperature $T$.
Magnetic energy is then approximately given as the initial energy on
scales $L(T)$. 
 
For fields which are maximally helical, i.e. $\He = \He_{\rm max} \approx 
\langle B^2(l)\, l\rangle \approx B^2(L)\, L$, 
Eq.~({\ref{relax}) may still be used to obtain the coherence scale of
  the field.  
Nevertheless,
due to helicity density $\He = (1/V)\int {\rm d}^3x\,\vA\cdot\vB$ (where
$\vA$ is the vector potential and $V$ is the integration volume)  
being an ideal invariant in the early Universe, a direct cascade of
energy from large scales to small scales is accompanied by an 
inverse cascade of energy from small scales to large scales.
That is, whereas in the absence of helicity the large-scale
field remains unprocessed, in the maximally helical case large-scale
fields undergo growth even on scales $l\simge L(T)$. Surprisingly, we find
that during this process of large length scale magnetic field 
amplification 
the initial spectral index $n$ is conserved~\cite{chris01}. 
The decay of magnetic energy, is thus described
by the requirement of conservation of helicity, in conjunction with an
increase of $L(T)$ described via Eq.~(\ref{relax}). 
Due to a vast increase of $L(T)$
in the early Universe, even initially sub-maximally helical fields,
i.e. $\He_{g} = h_g\He_{\rm max,g}$, with $h_g < 1$, 
ultimately reach a maximal helical configuration. Here, and throughout,
an index ``$g$'' denotes properties at the magnetogenesis epoch.
Parameterizing the
initial (comoving) magnetic energy spectrum 
by $\rho_B\approx\rho_{Bg}(l/L_g)^{-n}$, where $L_g$ is the initial
(magnetogenesis) coherence scale obeying Eq.~(\ref{relax}) 
and $\rho_{Bg}$ is approximate initial magnetic energy, one finds that fields
have reached a maximally helical state when $L(T)$ has grown beyond
\begin{equation}
L^{\rm max} \simeq L_gh_g^{-1/(n-1)}\, \label{max}. 
\end{equation}

This picture may be employed to derive damping exponents, 
i.e. $E_{\rm mag}\sim t^{-\gamma}$, and compare to those
inferred from numerical simulations (cf. Fig.~\ref{fig:E_mag_by_time}). 
Whereas the comparison is quite favorable in the viscous regime, turbulent
decay is observed somewhat slower than predicted. For example, for
non-helical, turbulent MHD with a $n=3$ spectrum we predict $\gamma =1.2$
whereas the best fit to the numerical simulation yields $\gamma = 1.05$.
Nevertheless, we argue that this trend, seen also by 
others~\cite{bisk99,chris01}, 
must be due to limited numerical resolution. Its explanation, i.e. 
a putative additional increase of relaxation time beyond $l/v$
with continuing evolution, requires the existence of additional dimensionful
quantities (of dimension velocity or length) associated with the flow.
As we may find none, other than $l_{\rm diss}$ (which is mostly 
widely separated from $L$ in the early Universe, but not in the
simulation) or $L_{\rm box}$, the size of the simulation box, we believe the
effect to be numerical in nature. We have noted, that spectra
at late times show a peak region $\Delta L$ quite spread, and are likely
only marginally resolved by the simulations. In any case, larger
numerical simulations are required to address this effect, with resulting
predictions for the surviving magnetic fields, given in this letter, 
being on the conservative side.

\bef
\def\eps@scaling{.96}
\showone{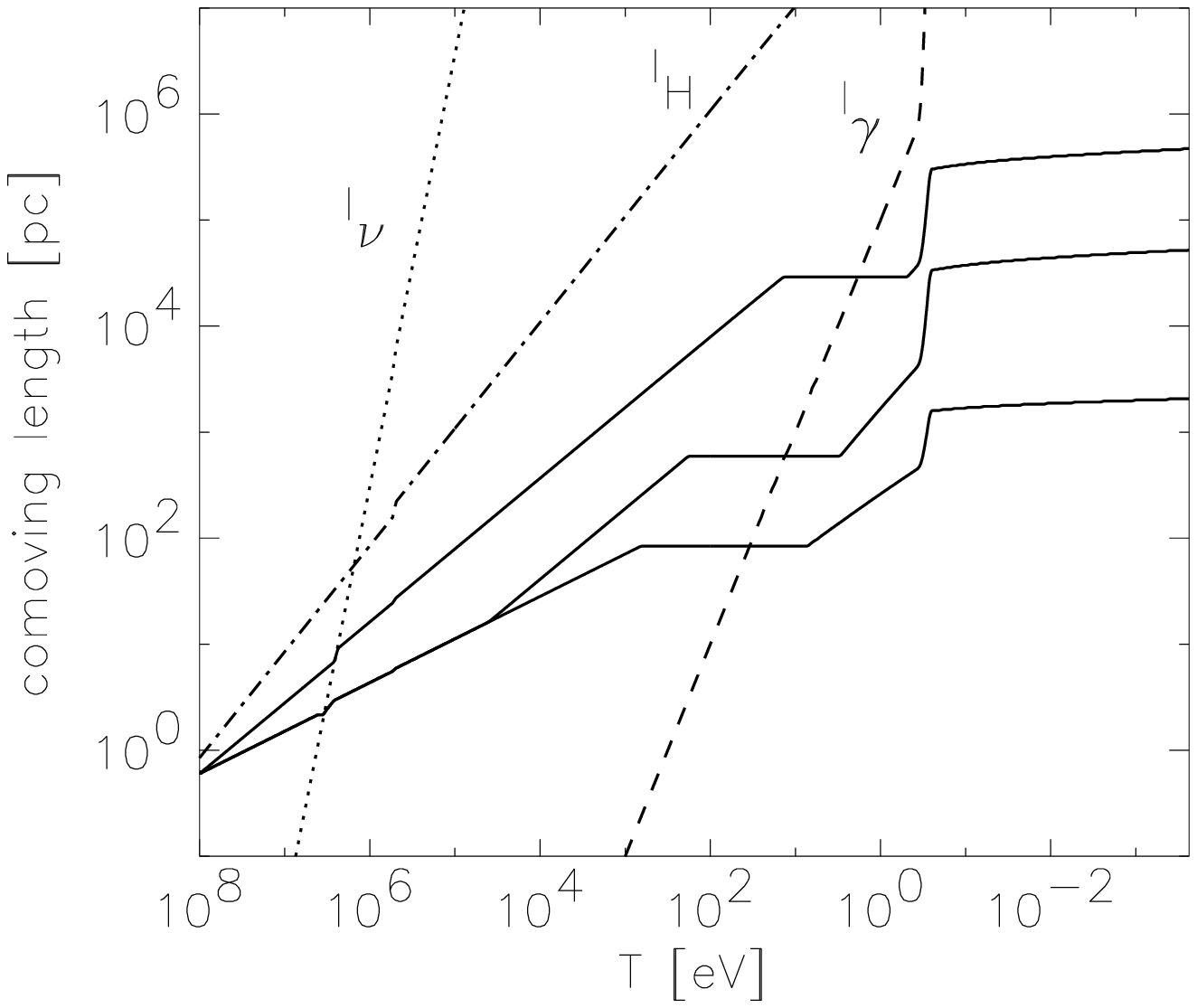}
\caption{Evolution of comoving coherence length for different initial
  magnetic field configurations with $r_g = 0.083$ and $n=3$ (solid lines from top
  to bottom: $h_g=1$, $h_g=10^{-3}$, $h_g=0$). The mean free paths of
  neutrinos and photons are labeled by $l_{\nu}$ and $l_{\gamma}$,
  and $l_H$ is the Hubble length.}
\label{fig:length_qcd}
\eef

We have undertaken the in practice straightforward but 
arduous effort to assemble these 
results and, under the inclusion of all appropriate dissipation terms,
and for quite general initial conditions, followed the growth of
magnetic coherence length and energy density from the very earliest times
to the present~\cite{BJ03}. Fig.~\ref{fig:length_qcd} shows examples
for the growth of $L(T)$ 
for a number of scenarios of magnetogenesis at the QCD phase transition. 
The evolution is observed as an
alternation between turbulent MHD and viscous MHD.
``Viscosity'' here is early on due to neutrinos, some time before
recombination due to photons, and after recombination due to hydrogen-ion
scattering and hydrogen-hydrogen scattering. Particularly notable are
phases where the growth of $L(T)$ is halted completely. This occurs either
at epochs before recombination
in the viscous regime with diffusing photons or neutrinos, as well in part
of the regime when those particles are free-streaming or at 
epochs after recombination, due to the
peculiar redshifting of Eq.~(\ref{relax}) 
and/or the effects of hydrogen diffusion
and ambipolar drag. Note, however, that the growth of $L(T)$ and concomitant
decrease of $B(T)$ during
the late phases of viscous MHD with free-streaming photons (neutrinos) 
may be faster than the growth of $L(T)$ during turbulent MHD.
Note also that initial conditions leading to relatively strong
magnetic fields at recombination result in a rapid increase of $L(T)$
at $T_{\rm rec}\approx 0.3\,\eV$, whereas for weaker fields 
$B\simle 10^{-13}\,\G$ a similar jump occurs at reionization. 
Note that effects due to structure formation are not taken into 
account here (see below, however).


We give here the final (pre-structure formation) results on the
coherence scale $L(T_0)$ and field amplitude $B(T_0)$, where $T_0$ is
present CMB temperature,
derived by employing
Eq.~(\ref{relax}), 
as well as retaining the initial field energy due to all scales 
$l\simge L$ in the submaximally helical case, and conserving
helicity density in the maximally helical case.
Complete results on $L$ and $B$ for all eras may be found 
in~\cite{BJ03}.
Fields which remain still submaximally helical 
(i.e. $L(T_0)\simle L^{\rm max}_c$)~\cite{remark0}
at the present epoch have for coherence length and field
strength

\bea
B(T_0) & \approx & 1.65\times 10^{-6}\,\G\,\, x^{n/(n+2)}
 \nonumber\\    
   & &      \times\left(\frac{r_g}{0.083}\right)^{1/2}
            \left(\frac{T_g}{100\,\MeV}\right)^{-{n/(n+2)}}
\\
L(T_0) & \approx & L_{gc}\, x^{-2/(n+2)} \, 
     \left(\frac{T_g}{100\,\MeV}\right)^{2/(n+2)} \quad,
     \label{lnonh}
\eea
where $x= 2.30\times 10^{-9}$ is a small factor and 
$r_g = (\rho_B/s_r^{4/3})_g$ is a convenient measure of magnetic energy
density $\rho_B$ in terms of radiation entropy density 
$s_r = (4/3)\,g_g\,(\pi^2/30)\,T_g^3$ at the 
magnetogenesis epoch assumed to occur at temperature $T_g$. Note that, 
somewhat optimistically $r_g = 0.083$, when magnetogenesis results in magnetic
energy density equivalent to the photon energy density shortly after a
QCD-transition with $g_g\approx 10.75$.  
The comoving coherence length $L_{gc}$ at the magnetogenesis
epoch is given by
\begin{equation}
L_{gc} \approx 0.45\,\pc\,\sqrt{n}
            \left(\frac{r_g}{0.083}\right)^{1/2} \,
            \left(\frac{T_g}{100\,\MeV}\right)^{-1} \, .
\end{equation}
This yields, for example, for $r_g=0.083$, $T_g=100\,\MeV$, $n=3$ to the
appreciable field strength of $B_c\approx 1.1\times 10^{-11}\,\G$.
If we were to apply the simulation observed instead of theoretically predicted
decay exponent $\gamma$ the result would increase approximately to $B\sim
5\times 10^{-11}\,\G$.
On the other hand, fields which have reached a maximally helical 
(i.e. $L(T_0)\simge L^{\rm max}_c$) state at present have

\bea
B_c & \approx & 4.69\times 10^{-12}\,\G\,\, y \,
\\
L_c  & \approx & 550\, \pc\,\, 
   y\,\sqrt{n}\, \label{lhel}
\eea
with
\begin{equation}
y = \left(\frac{r_g}{0.083}\right)^{1/2} \,
    \left(\frac{h_g}{10^{-8}}\right)^{1/3} \,
    \left(\frac{T_g}{100\,\MeV}\right)^{-1/3}\nonumber
\end{equation}
where $h_g$ is the fractional helicity of the maximal one $\He_{{\rm max},g}$
at the generation epoch.


\bef
\showone{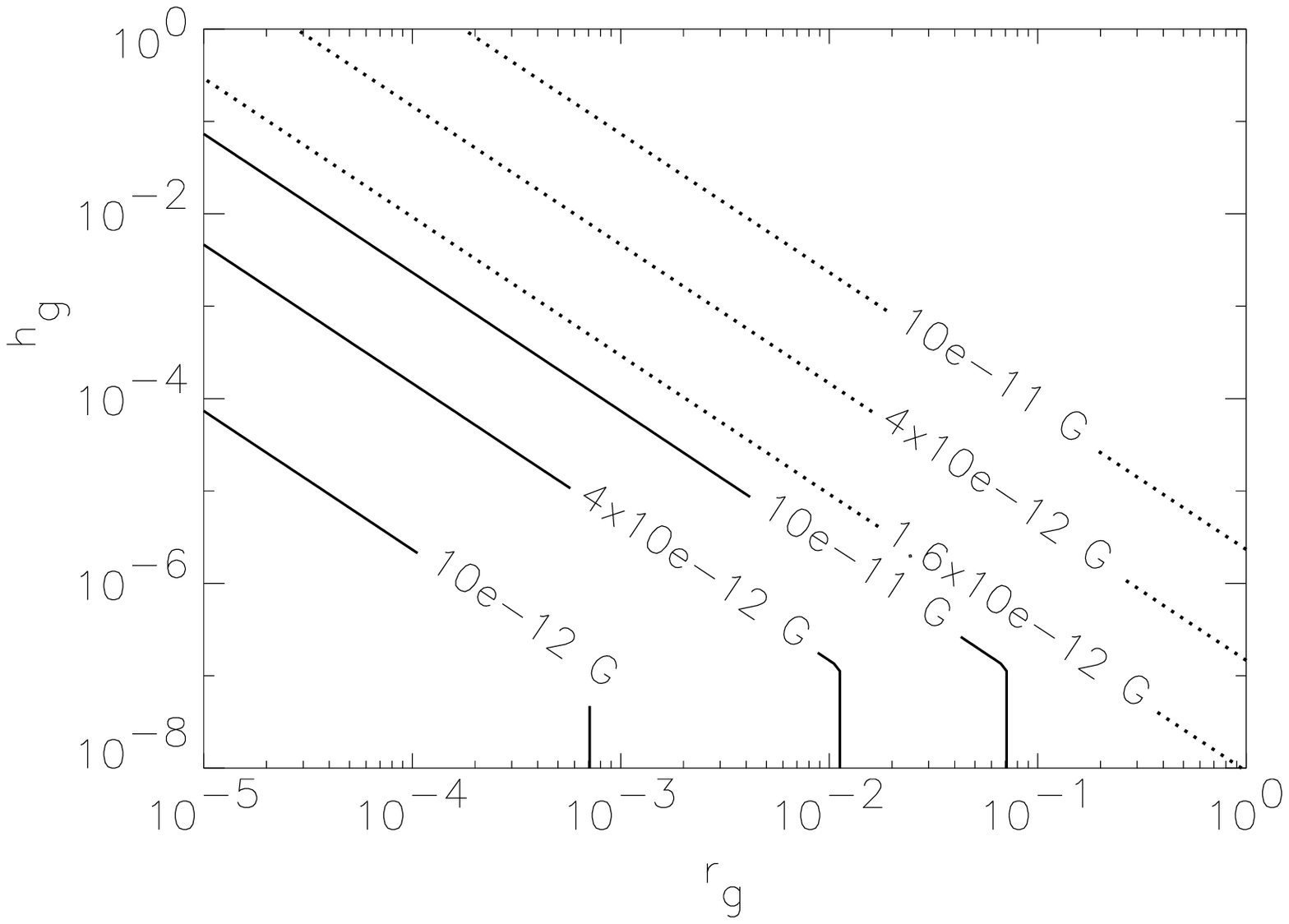}
\caption{Final magnetic field strengths in the $r_g - h_g$ parameter
  space for $T_g = 100\,\MeV$ (solid lines) and $T_g = 100\,\GeV$
  (dotted lines), where we assumed a spectral index of $n=3$.}
\label{fig:contour}
\eef

Though on small scales, it is seen that surprisingly strong fields may
survive an early Universe magnetogenesis epoch to the present
(cf. Fig.~\ref{fig:contour}). 
This is interesting in light of recent simulations on the formation of
clusters of galaxies from slightly overdense and pre-magnetized regions via
the gravitational instability~\cite{DBL99}. The authors find, that fields
of strength $B_c\approx 4\times 10^{-12}\,\G$ (corresponding to their
quoted $B\approx 10^{-9}\,\G$ at simulation starting redshift $z = 15$)
yield to clusters whose Faraday rotation measures are essentially 
indistinguishable from those observed in real clusters~\cite{remark1}. 
Furthermore, the authors arrive at the intriguing conclusion, that this
result is virtually independent of whether a homogeneous field is 
assumed initially, or a field whose energy contribution is dominated
by fluctuations on the 
very smallest scales in their simulations~\cite{remark2}. 
It is not clear if
such an erasure of memory of initial conditions, possibly related to an
interplay between the development of shear flows and small-scale turbulence
during the course of gravitational collapse, pertains if initial field
coherence lengths in the cluster simulations are reduced by a further 
factor $\sim 100$
(due to the comparison between typical coherence lengths in 
Eqs.~(\ref{lnonh}), (\ref{lhel})
and the spatial resolution, $\sim 100\,\kpc$ comoving, of the simulations).
Nevertheless, if so, cosmological magnetic fields generated during early
eras, either of moderate magnetic helicity, or generated fairly late, 
could account for present-day observed cluster magnetic fields, and as such
in the absence of any further dynamo amplification. This holds true even
if as recently claimed such fields have very blue sprectra~\cite{DC03}.
We conclude that this
interesting possibility seems to deserve further investigation.

\vskip 0.1in
We acknowledge invaluable support and enthusiasm by T. Abel, A. Kercek, and
M. Mac Low in the numerical realization of this study.


\begin{thebibliography}{99}
\bibitem{remark1} Fields of this final strength may also be
shown to be too weak at redshift $z\approx 10^6$ to induce significant
spectral distortions in the cosmic microwave background 
radiation~\cite{JKO00}
\bibitem{remark2}
It is important to note that employed initial magnetic field spectra
in Ref.~\cite{DBL99}, $P(k)\sim k^{-5/3}$, correspond to a blue
spectral index $n=4/3$ in our notation, 
as opposed to Kolmogoroff $n=-2/3$~\cite{D03}.
\end{thebibliography}
\end{document}